# Anomalous itinerant magnetism in single crystal $Sr_4Ru_3O_{10}$: A thermodynamic and transport investigation

G. Cao, S. Chikara and J. W. Brill

Department of Physics and Astronomy, University of Kentucky, Lexington, KY 40506

P. Schlottmann

Department of Physics, Florida State University, Tallahassee, FL 32306

Abstract

A thermodynamic and transport study of $Sr_4Ru_3O_{10}$ as a function of temperature and magnetic field is presented. The central results include a growing specific heat C with increasing field B, a magnetic contribution to C/T at low temperatures proportional to –log(T), an abrupt jump and a peak in C/T at 2.90 T and 7 T for B||ab-plane and B||c-axis, respectively, and corresponding changes in the low T power laws of the resistivity. The novelty of this work lies in the fact that this system is strongly anisotropic displaying spontaneous ferromagnetism along the c-axis and an intralayer metamagnetic transition with a possibility of a nearby quantum critical point. The exotic behavior reflects new physics that is yet to be understood.

## I. Introduction

A quantum critical point (QCP) can be realized by tuning the critical temperature of a phase transition to absolute zero by varying external parameters such as the magnetic field, pressure or the doping level. While a phase transition at finite T is governed by thermal fluctuations, the relevant fluctuations close to a QCP are of quantum mechanical nature. A QCP usually impacts the physical properties over a wide range of temperatures. Quantum criticality is well illustrated in heavy fermions [1] and high $T_c$ cuprates, and, recently, in materials intimately associated with itinerant metamagnetism. For instance, $UPt_3$ [2], $CeRu_2Si_2$ [3] and $Sr_3Ru_2O_7$ [4] are enhanced paramagnets showing a field-induced metamagnetic transition. When the critical end point of the metamagnetic transition is tuned to zero temperature, the critical fluctuations cause the breakdown of Fermi liquid behavior with profound consequences, such as divergent specific heat [C/T ~ -log(T)] and unusual power laws in the resistivity at low T [1-6]. However, unusual power laws are also observed when the critical temperature remains finite. In this paper, we report the temperature and field dependence of the specific heat, resistivity and magnetization of triple-layered $Sr_4Ru_3O_{10}$. Critically driven by both, intralayer field-induced metamagnetism and interlayer spontaneous ferromagnetism [7-12], unusual power laws in T are found. These features are distinct from those observed in other systems involving a QCP.

Our results for $Sr_4Ru_3O_{10}$ reveal a host of anomalous properties, namely, a growing specific heat C with increasing magnetic field B, a –log(T) contribution to C/T at low T, an abrupt jump and anomaly in C/T at $B_c$=2.90 T and $B_c$=7 T for B||ab-plane and B||c-axis, respectively, and corresponding changes in the power law of the resistivity

between $T^2$ and $T^{3/2}$. The novelty of this work lies in the fact that the quantum fluctuations occur in a system where both interlayer spontaneous ferromagnetism and intralayer field-induced metamagnetism coexist, a feature distinctively different from all other systems involving a QCP. (The term "spontaneous ferromagnetism" is to distinguish it from field-induced ferromagnetic order or metamagnetism because this system exhibits both.)

$Sr_4Ru_3O_{10}$ belongs to the layered Ruddlesden-Popper series, $Sr_{n+1}Ru_nO_{3n+1}$ (n=number of Ru-O layers/unit cell). The comparable and competing energies of crystalline fields (CEF), Hund's rule interactions, spin-orbit coupling and electron-lattice coupling crucially determine the CEF level splitting and the band structure, hence the ground state. The physical properties are thus highly dimensionality (or n) dependent. As a result, the magnetic state of $Sr_{n+1}Ru_nO_{3n+1}$ systematically evolves from paramagnetism (n=1), enhanced paramagnetism (n=2) to spontaneous ferromagnetism (n=∞) with increasing n. Situated between n=2 and n=∞, $Sr_4Ru_3O_{10}$ (n=3) displays interesting phenomena ranging from quantum oscillations [8], tunneling magnetoresistance [8], unusual low temperature specific heat [9], strong spin-lattice coupling [10, 11], switching behavior [12] and other properties [13,14]. The unique feature, however, is borderline magnetism: For a field *along the c-axis* (perpendicular to the layers), $Sr_4Ru_3O_{10}$ displays spontaneous ferromagnetism at a Curie temperature, while for a field within *the ab-plane* it features a pronounced peak in magnetization and a first-order metamagnetic transition [7-11]. The c-axis ferromagnetism indicates that the Stoner criterion is satisfied, $Ug(E_F)\geq 1$, where U is an exchange interaction and $g(E_F)$ the density of states at the Fermi surface. Below $T_C$ the spin-up and spin-down bands are thus spontaneously split by the exchange splitting Δ in the absence of an applied field. The response to a field in the plane, however, is strikingly similar to Stoner enhanced paramagnetism with $Ug(E_F)<1$ [4]. The coexistence of the interlayer ferromagnetism and the intralayer metamagnetism makes $Sr_4Ru_3O_{10}$ significantly different from $Sr_3Ru_2O_7$ where spontaneous ferromagnetism is absent.

The crucial properties of $Sr_4Ru_3O_{10}$ discovered from previous studies [7-11] are the following: (1) The c-axis magnetization, $M_C$, is ferromagnetic with $T_C$ at 105 K followed by an increased spin polarization below $T_M$=60 K with large irreversibility upon in-field and zero-field cooling [7-8]. In contrast, the ab-plane magnetization, $M_{ab}$, is much smaller and exhibits a weak cusp at $T_C$ and a broad peak at $T_M$ [7-8]. (2) The isothermal magnetization $M_C$ illustrates that the spins are readily polarized and saturated along the c-axis at B=0.2 T, yielding a saturation moment of 1.2 $\mu_B$/Ru. On the contrary, $M_{ab}$ displays a first-order metamagnetic transition at $B_c$ [8]. (3) Our early transport study has showed a sign of a divergent coefficient A of $\rho \sim AT^2$ (where ρ is the resistivity) near metamagnetic transition $B_c$ [8]. (4) The Raman study [10] demonstrates a distinct structural contribution to the metamagnetic transition that is directly associated with $RuO_6$ octahedral rotations. Specifically, the negative derivative of the $B_{1g}$ phonon frequency with respect to the application of an in-plane field suggests that the $RuO_6$ octahedra are unstable to distortions that increase the in-plane RuO bonds upon the application of the an in-plane field. (5) The neutron study has found no antiferromagnetic Bragg peaks [14], implying the abrupt jump in M(B) is not a simple spin-reorientation transition from antiferromagnetic to ferromagnetic state. (6) The temperature dependent (B=0) specific heat of the present sample is the same as that previously reported for a sample measured with ac-calorimetry at B=0 [9]. The early heat capacity study has found that an approximate lower limit entropy change $\Delta S > 0.02R$ near $T_C$ (where $R = 8.314\ J\ mol^{-1}\ K^{-1}$ is the gas constant) and an upper limit $\Delta S \sim \Delta C \sim 0.4R$, as expected for a mean-field anomaly [9]. However, if all three spins/formula unit order, the expected entropy change is $\Delta S = 3.3\ R$, thus the measured entropy change is an order of magnitude smaller than expected for complete spin ordering, suggesting that either the spin ordering is not spatially uniform or that only a small component of the spins order. It is noted that the magnetic entropy removal at $T_C$ is generally small for weakly ferromagnetic metals, chiefly due to spin fluctuations. Furthermore, low temperature C/T vs $T^2$ (B=0) plot exhibits a negative curvature for T < 9 K [9]. Such curvature implies that, in addition to the usual phonon ($C_{ph} = \beta T^3$) and electronic ($C_e = \gamma T$) contributions to the specific heat, there must be a term $C_S \sim \eta T^p$ with $1<p<3$, presumably due to spin excitations. For the fits to give (per atom) Debye temperatures below ~500 K, we must have $p < 2$. (If one estimates $\beta$ from the linear portions of the curves only, i.e. from the data for T > 9 K, one obtains an average Debye temperature for the two samples of $\Theta = (367 \pm 15)$ K, close to the values for cubic (i.e. $n = \infty$) $SrRuO_3$ and double layer $Sr_3Ru_2O_7$. Therefore, including a term with $p<3$ decreases the value of $\beta$ and increases $\Theta$.) These constraints on p suggest magnetic order in apparent contradiction to the measured magnetic properties, again indicating the complexity of the magnetic order [9]. Finally, our early study has already mentioned that the low temperature specific heat can also be fit with a spin term $C_S = \eta T \ln(T/T_0)$, as for ferromagnetic fluctuations near at quantum critical point [15]. While it is difficult to interpret the temperature dependence of heat capacity, the complex behavior presented clearly points to a possible exotic state where there exists a delicate balance between order and fluctuations [9]. It is in part this complex behavior that has motivated the current study presented in this paper.

**II. Experimental**

The single crystals studied were grown using flux techniques. All single crystals were grown in Pt crucibles from off-stoichiometric quantities of $RuO_2$, $SrCO_3$ and $SrCl_2$ mixtures with $SrCl_2$ being self flux. The mixtures were first heated to 1480 °C in a Pt crucible covered by a Pt cover, soaked for 25 hours, and slowly cooled at 2-3 $^oC$/hour to 1380 $^oC$ and finally cooled to room temperature at 100$^oC$/hour. The starting Sr:Ru ratio and the thermal treatments are critical and subtle for the formation of perovskite crystals as nucleation of its sister compounds $SrRuO_3$ and $Sr_2RuO_4$ are also energetically favorable. By carefully changing the ratio and thermal treatments, we have successfully grown crystals of $Sr_{n+1}Ru_nO_{3n+1}$ with n=1, 2, 3 and ∞. These crystals are characterized by single crystal x-ray diffraction, EDX and TEM [8]. All results suggest that the crystals studied are of high quality with *no impurity phases and no intergrowth*. The high quality of our samples is further confirmed by the observation of quantum oscillations and a small Dingle temperature of 3 K, a measure of impurity scattering [8]. It needs to be pointed out that we have grown the $Sr_4Ru_3O_{10}$ crystals using *both the floating zone and flux techniques* and comparatively studied these crystals grown from the two different techniques. The thermodynamic results of this comparative study were reported earlier [9]**.** *It is now well recognized that the floating zone technique cannot yield pure* $Sr_4Ru_3O_{10}$ *crystals without the inclusion of a significant amount of second phase such as the strong ferromagnet* $SrRuO_3$ *whose Curie temperature is at 165 K. This unwanted phase very often overshadows the intrinsic properties of* $Sr_4Ru_3O_{10}$ *or yields misleading* *results. An obvious indication of the inclusion of the impurity $SrRuO_3$ is a pronounced anomaly near 165 K, the Curie temperature of $SrRuO_3$, in temperature dependence of magnetization, a rather weak and broad metamagnetic transition in isothermal magnetization, and a particularly small or negligible heat capacity change near $T_C$=105 K. These characteristics of the impurity inclusion are seen in some published literatures.* Heat capacity measurements were performed on a 2.7 mg single crystal using a Quantum Design PPMS that utilizes a thermal-relaxation calorimeter operating in fields up to 9 T. Magnetic and transport properties were measured using a QD 7T magnetometer and a 15T Oxford magnet.

**III. Results and Discussion**

The application of B tends to align spins and usually suppresses spin fluctuations and hence the specific heat. It is striking that the specific heat of $Sr_4Ru_3O_{10}$ responds to B oppositely. Fig.1 shows the specific heat divided by T as a function of T for $1.8\leq T\leq 12$ K with B||ab-plane (a) and c-axis (b), respectively. Data for T>12 K is not shown for clarity. There are a few crucial features. Firstly, C/T at low T increases radically for B||ab-plane, particularly in the vicinity of the metamagnetic transition (Fig.1a) in contradiction to the anticipated behavior, implying a considerable enhancement of the quasi-particle m* in $g(E_F)$. Secondly, C/T for B||c-axis shows vastly different T-dependence (Fig.1b), decreasing with increasing B for T>6 K, consistent with a suppression of spin fluctuations as anticipated for a regular magnetic state. But it grows, though less drastically, for T<6 K by showing a broad peak near 8 T, suggesting that unexpected low-energy excitations develop in the spontaneous ferromagnetic state. Thirdly, for both B||ab-plane and B||c-axis, C/T for $B\leq 2.7$ T and T < 10 K nearly follows a linear T-

dependence, $C/T \sim a+bT$. The rapid increase of $a$ with B could be indicative of a Fermi liquid with a nearby 2D critical point [6]. This interpretation requires a very small phonon contribution, $\beta T^2$ (apparent in the positive curvature in C/T vs. T for T > 10 K) and therefore an unphysically large value of the Debye temperature. Alternatively, as discussed in [9], the specific heat can be fit by $C/T = \gamma + \beta T^2 + \delta T^{1/2}$ (with a typical value of $\beta \sim$ 0.04 mJ $mol^{-1}$ $K^{-4}$), where the $T^{1/2}$ term is associated with an unexpectedly large contribution from ferromagnetic spin-waves. As B is increased further the T-dependence of C/T weakens from almost T-linear at $B<B_c$ to a shoulder with a small peak at B=5 T and a plateau for $B \geq 6T$ (Fig.1). To further emphasize the unusual temperature dependence of C/T in fields, C/T vs. $T^2$ is plotted in Fig.1c, where the negative curvature associated with the $T^{1/2}$ term is manifested for T< 8 K [9]. The inset clearly shows the sharp peak at 2.5 K for B=5 T that diminishes at higher fields and eventually evolves into a rapid downturn at 9 T.

It is clear that the amplitude of fluctuations rapidly grows as B increases. To separate the field-induced contribution ΔC from other contributions to C, we subtract the zero-field C(0) from the in-field C(B), i.e., ΔC=C(B)-C(0). ΔC/T plotted as a function of T is shown in Fig.2 for B||ab-plane (Fig.2a) and B||c-axis (Fig.2b). A dominant feature is that ΔC/T increases logarithmically with decreasing T as B rises. The slope of the log-dependence gets stronger when B approaches 5 T and 7 T for B||ab-plane and B||c-axis, respectively. (While there is a gradual change in slope (at ~ 4K) in the log plot (see inset), the variation of ΔC/T is approximately logarithmic for a range of over 10 K. Furthermore, for T < 4K, the slopes are constant where ΔC/T changes by a factor of 2.) At 1.8 K, ΔC/T for B||ab-plane increased by a factor of three from less than 0.04 J/mol $K^2$ at 1 T to nearly

0.12 J/mol $K^2$ at 5 T. At higher T, there exists a broad peak marked by an arrow. It moves to lower T and eventually vanishes as B increases so that the entropy is shifted into the logarithmic upturn. Remarkably, for B||ab-plane, as B approaches $B_c$, C/T shows a jump at B=2.8 T and near 10 K. This jump develops into a pronounced peak at $B_c$=2.9 T and near 7 K and then broadens for $B>B_c$. For B||c-axis, no such an abrupt jump in C/T is seen near $B_c$. The broad peak similar to that for B||ab-plane is progressively suppressed by B and vanishes at B>6 T.

The abrupt jump in C/T and the metamagnetic transition for B||ab-plane are once more emphasized in Fig.3, in which the field dependence of C/T is compared to that of the resistivity and magnetization. The jump in C/T is drastic and persists up to ~12 K, suggesting a critical end point of the metamagnetic transition (Fig.3b). Relevantly, La-doping pushes the anomalies to lower T and B, indicating that a QCP is reached near 13% of La-doping [11]. As shown in Fig.3c (left scale) the metamagnetic transition also affects the magnetoresistence ratio, defined as $\Delta\rho_c/\rho_c(0)=[\rho_c(B)-\rho_c(0)]/\rho_c(0)$, which changes by more than 40% near $B_c$, confirming large spin fluctuations in a state without long-range order immediately above the transition.

The presence of the quantum fluctuations is further corroborated by the temperature dependence of the resistivity. Shown in Fig.4a is the ab-plane resistivity $\rho_{ab}$ at a few representative fields as a function of $T^2$ or $T^{5/3}$ (upper axis) for 1.7<T<17 K. The linearity of $\rho_{ab}$ suggests well-defined power laws followed at various B. Fig.4b maps the details of the coefficients of the $T^2$- and $T^{5/3}$-dependences of $\rho_{ab}$ as a function of B. $\rho_{ab}$ fits well to $\rho_{ab}=\rho_o+A_{ab}T^2$ for B<2 T where $\rho_o$, the residual resistivity, is 6 μΩ cm at B=0, indicative of the high purity of the sample. The coefficient $A_{ab}$, which depends on the effective mass m*, rapidly increases with B, indicating a divergent m*. $\rho_{ab}$ starts to deviate from the $T^2$-dependence at B=2.2 T, signaling the breakdown of Fermi liquid properties. For 2.2≤B<5.5 T $\rho_{ab}$ is proportional to $A_{ab}*T^{\alpha}$, where $\alpha$ is smaller than 2. $\alpha$ briefly varies between 1.5 and 1.6 for 2.0<B<2.4 T and then settles at 5/3 for 2.4≤B≤5.5T. The coefficient $A_{ab}*$ for the $\alpha$=5/3 dependence rises steeply with B, peaks at 2.9 T and decreases for larger fields. The Fermi liquid behavior is recovered for B>5.5 T. This is consistent with the behavior of C which at low T starts to decrease when B>5 T. $\rho_o$ consistently shows a similar field dependence and diverges near $B_c$ [8].

A singular T-dependence of the resistivity with $\alpha$< 2, specifically, $\alpha$=3/2 and 5/3 is seen in systems with quantum criticality such as MnSi [16], $Sr_3Ru_2O_7$ [4], heavy fermion systems [5] as well as impurity doped $Sr_4Ru_3O_{10}$ [11]. The $T^{5/3}$-dependence of resistivity is often attributed to either dominating low-angle electron scattering (low-*q* fluctuations) or high-*q* fluctuations scattering electrons in the vicinity of a QCP [16-18], hence weakening the temperature dependence from $T^2$. The power-law $T^{3/2}$ is thought to be associated with effects of diffusive motion of the electrons caused by the interactions between the itinerant electrons and critically damped magnons [19]. The change of $A_{ab}$ and $A_{ab}*$ with B entirely tracks C/T and $M_{ab}$, suggesting a possible proximity of a QCP and an intimate connection between the critical fluctuations and the metamagnetism.

What is equally intriguing is that C/T for B||c-axis, where the spontaneous ferromagnestism occurs, shows a weaker and yet well-defined peak at 7 T followed by a minimum at $B_m(c)$=8 T (Fig.3a). Unlike the jump in C/T for B||ab-plane, this peak sensitively changes with T, and nearly vanishes for T>6.8 K, where C/T no longer increases with B, the behavior expected for a regular metal. The anomaly in C/T also

accompanies a varying $\alpha$ of resistivity between 3/2 and 5/3 near 7 T for B||c-axis (not shown), and an abrupt change in slope of $\rho_c$ at $B_m(c)$=8 T (Fig.3c, left scale), suggesting a significant change in the scattering rate.

**IV. Conclusions**

The exotic behavior brings up interesting questions: Why do the spontaneously spin-split bands undergo the critical fluctuations? Is it possible that the metamagnetic critical fluctuations essentially arise from field-induced spin-split Fermi surfaces that are not well defined? If so, the data suggest that $g(E_F)$ or m* and long range correlations diverge at the critical point, leading to non-Fermi liquid behavior. It is intriguing that field-driven quantum critical fluctuations and nesting properties should not be expected given the fact that the c-axis ferromagnetism in $Sr_4Ru_3O_{10}$ is robust [7-11], thus the spontaneously spin-split bands are stable and well defined.

Long-range ferromagnetic order and a metamagnetic transition have been predicted within a simple Stoner model with a two-dimensional density of states, which has a logarithmically divergent Van Hove singularity [20]. The ruthenates are layered compounds and strongly anisotropic and, hence, 4d-bands with 2D-dispersions are a reasonable assumption. Only the three $t_{2g}$-orbitals play a role. With increasing n the series $Sr_{n+1}Ru_nO_{3n+1}$ is gradually driven from a paramagnet for n=1, to a metamagnetic paramagnet with QCP at n=2, to a ferromagnet with metamagnetic transition by the Fermi level approaching the Van Hove singularity [20]. For n=1 and 2 this is confirmed by band structure calculations [21, 22]. It is also consistent with La-doping for this compound [11]. Each trivalent La replacing divalent Sr provides an additional electron shifting the Fermi level. This scenario, however, is too simple to explain the magnetic anisotropy, which requires a coupling of the spins to the orbits. The $t_{2g}$-orbitals are split in the band structure due to the crystalline electric fields, so the xy-orbital (contained in the ab-plane) is favored. As a consequence of this splitting orbital order of some kind is expected for this compound. Together with the Hund's rule couplings (maximizing the spin and the spin-obit coupling) the spin and the orbital states are coupled perturbatively, favoring (in this case) the c-axis as the easy axis of magnetization. The ordered spins are believed to be slightly canted for $T_C < T < T_M$ and below $T_M$ the canting is either reduced or completely suppressed. At the metamagnetic transition for the field in the ab-plane then corresponds to a reorientation of the spins into the plane.

As seen in Fig. 3b the critical end-point of the first order metamagnetic transition is roughly at 13 K. La-doping pushes the anomalies to lower T and lower fields. This is an indication that in the range $0.13 < x < 0.15$ of La-doping a QCP is reached [11]. Ca-doping, on the other hand, introduces only minor changes, so that chemical pressure does not appear to be a tuning parameter for the QCP [11]. The effects of a QCP are often far-reaching and the results discovered in this paper [e.g., $\Delta C/T \sim -\ln(T)$ and power laws in the resistivity] are believed to be in part due to the quantum critical fluctuations. Nevertheless, this system is indeed unique and the exotic behavior it displays reflects new physics that is yet to be understood.

**Acknowledgements**: This work was supported by NSF grants DMR-0240813, DMR-0400938, DMR-0552267 and DOE grant DE-FG02-98ER45707.

**Figure Captions:**

**Fig.1**. Specific heat divided by temperature, C/T, as a function of T for (a) B||ab-plane and (b) B||c-axis; (c) C/T for B||ab-plane as a function of $T^2$ for $0 \leq B \leq 5$ T; Inset 1: C/T for B||a-plane vs T for a few representative fields; Inset 2: C/T for B||ab-plane vs. $T^2$ for $5 \leq B \leq 9$ T.

**Fig.2**. Logarithmic temperature dependence of [C(B)-C(0)]/T for (a) B||ab-plane and $0 \leq B \leq 5$ T (inset: for $5 \leq B \leq 9$ T), and for (b)B||c-axis. The arrows track the shift of the peak.

**Fig.3**. Field dependence of C/T for (a) B||c-axis and (b) B||ab-plane for various temperatures; (c) $[\rho_c(B)-\rho_c(0)]/\rho_c(0)$ (current along the c-axis) (left scale) at T=0.6 K and isothermal magnetization M for B||ab-plane and B||c-axis (right scale) at T=1.7 K.

**Fig.4.** (a) The ab-plane resistivity $\rho_{ab}$ at a few representative fields as a function of $T^2$ (lower axis) or $T^{5/3}$ (upper axis) for a range of 1.7<T<17 K; (b) The field dependence of the coefficient $A_{ab}(\alpha=2)$ (left scale) and $A^*_{ab}(\alpha=5/3)$ (right scale) for B||ab-plane.

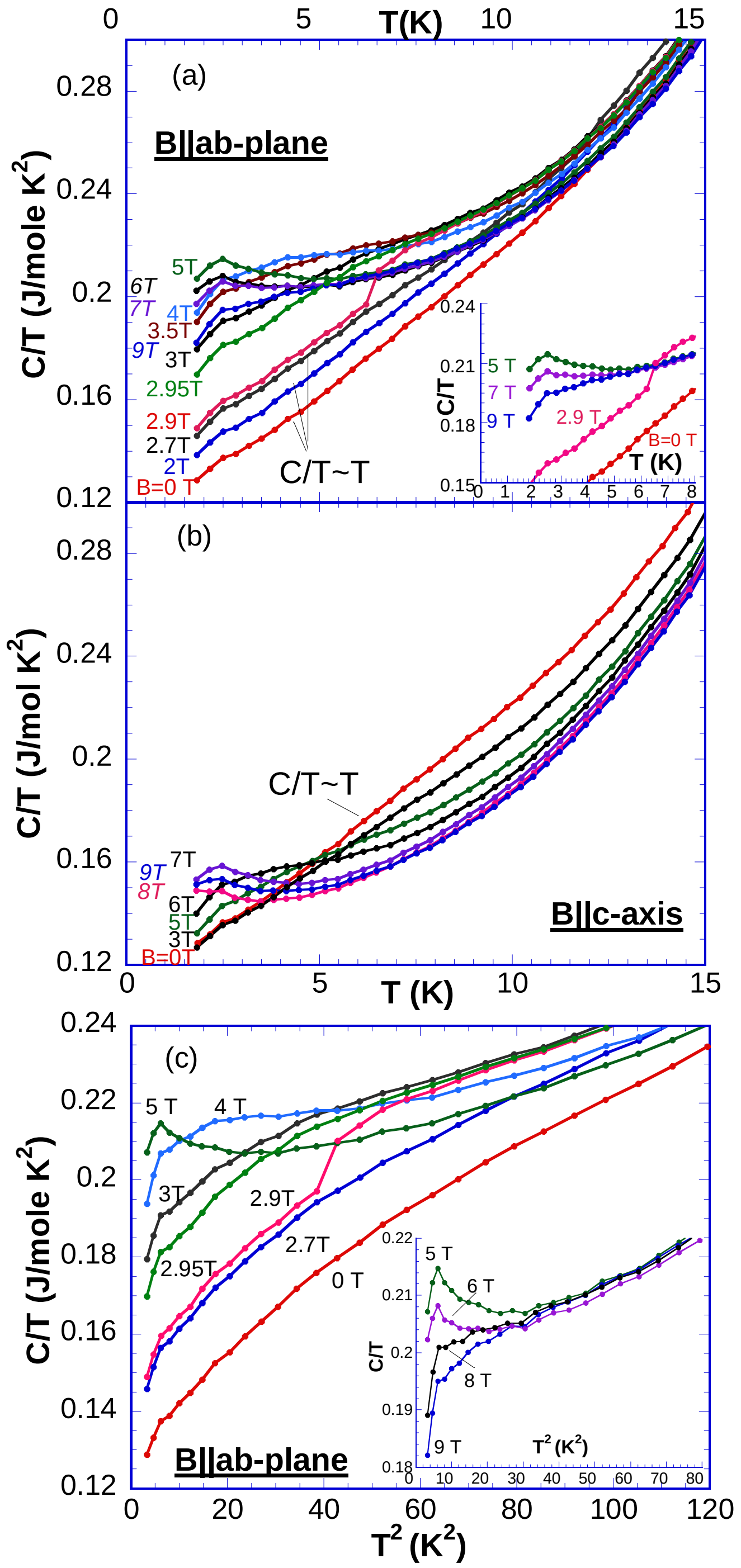

Fig.1

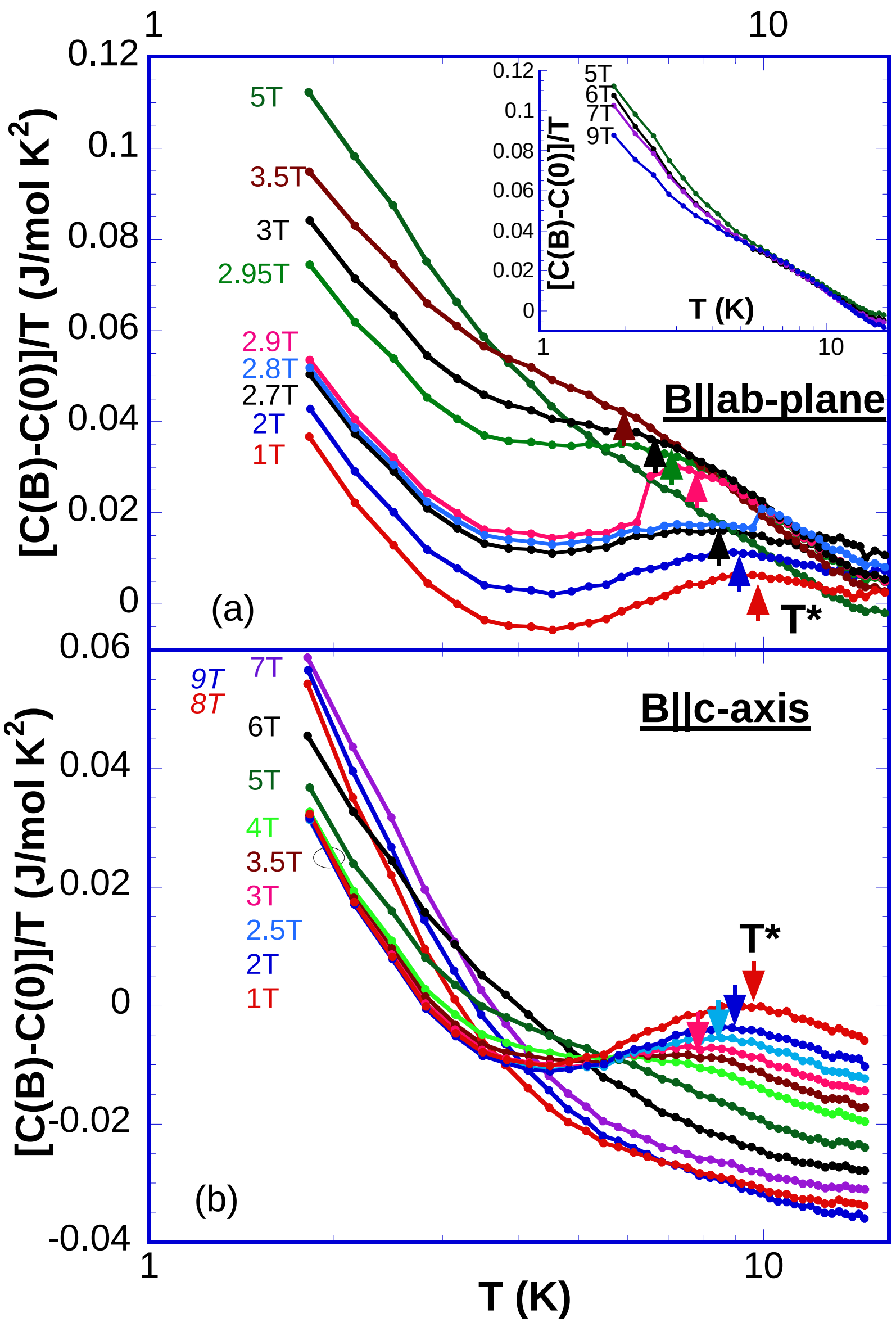

Fig.2

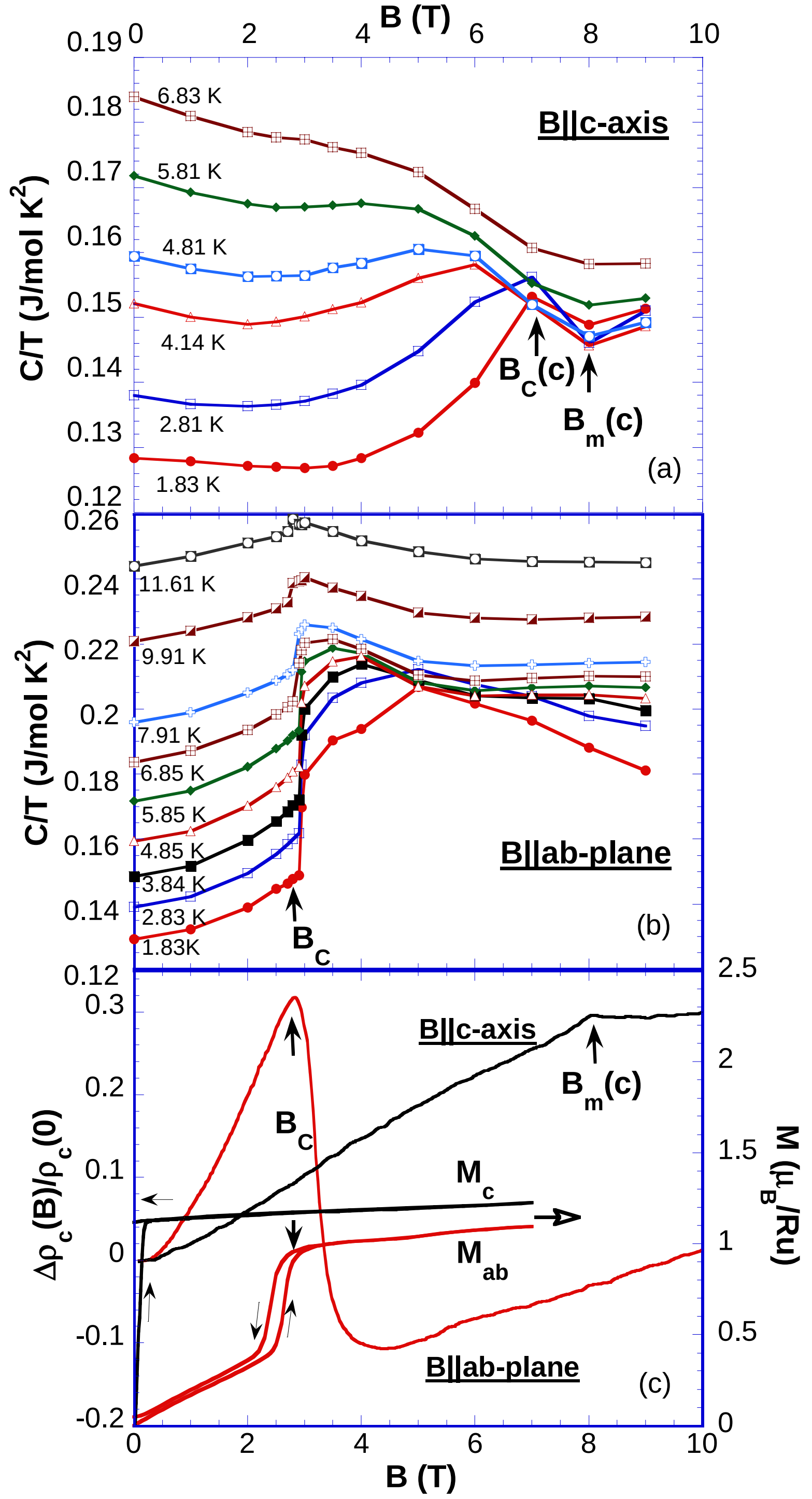

Fig. 3

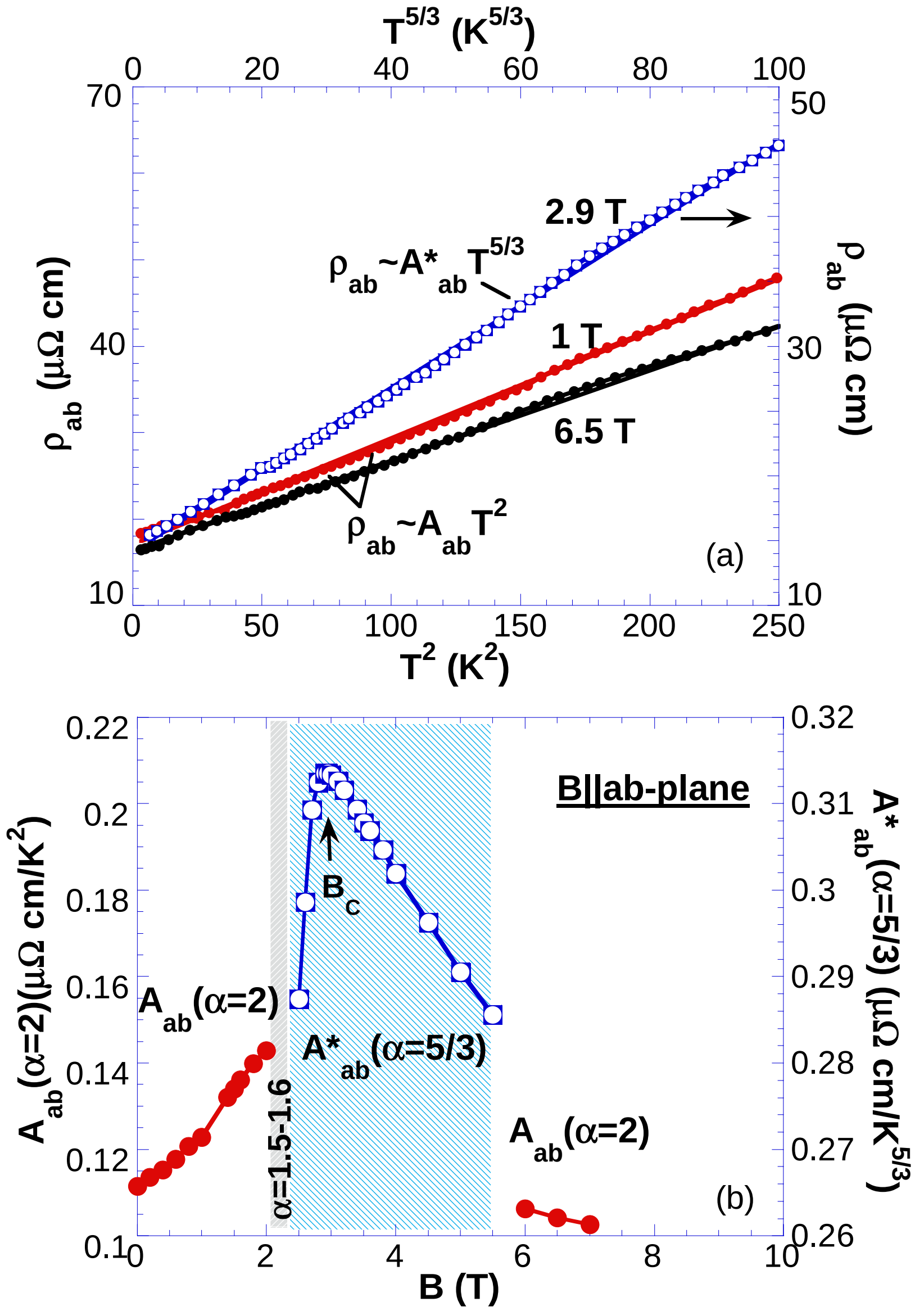

Fig. 4